\documentclass[pmlr]{jmlr}

\usepackage{bm}
\usepackage{mathtools}
\usepackage{booktabs}
\usepackage{enumitem}
\usepackage{subcaption}
\usepackage{microtype}
\usepackage{multirow}
\makeatletter
\g@addto@macro\normalsize{%
  \setlength{\abovedisplayskip}{5pt plus 1pt minus 2pt}%
  \setlength{\belowdisplayskip}{5pt plus 1pt minus 2pt}%
  \setlength{\abovedisplayshortskip}{2pt plus 1pt}%
  \setlength{\belowdisplayshortskip}{3pt plus 1pt minus 1pt}%
}
\makeatother

\newif\ifsupp\supptrue

\jmlrvolume{vvv}
\jmlryear{2027}
\jmlrworkshop{9th Annual Conference on Learning for Dynamics and Control}

\newcommand{\R}{\mathbb{R}}
\newcommand{\E}{\mathbb{E}}
\newcommand{\Prob}{\mathbb{P}}
\newcommand{\T}{\mathsf{T}}
\newcommand{\trace}{\operatorname{tr}}
\newcommand{\mc}{\mathcal}
\newcommand{\ind}{\mathbb{I}}
\newcommand{\ones}{\mathbf{1}}
\DeclareMathOperator{\col}{col}

\DeclareMathOperator{\blkdiag}{blkdiag}
\DeclareMathOperator{\diag}{diag}
\newcommand{\Ed}{\E_{\delta}}
\newcommand{\norm}[1]{\lVert #1\rVert}
\newcommand{\sm}{\sigma_{\min}}
\newcommand{\lmin}{\lambda_{\min}}
\newcommand{\lmax}{\lambda_{\max}}
\newcommand{\eid}{\epsilon^{\mathrm{id}}}
\newcommand{\ew}{\epsilon^{\mathrm{w}}}
\newcommand{\eb}{\epsilon^{\mathrm{b}}}
\newcommand{\epsm}{\epsilon^{\mathrm{m}}}
\newcommand{\ef}{\epsilon^{\mathrm{f}}}

\newcommand{\Ai}{\hat A^{(i)}}
\newcommand{\tth}{\tilde\theta}
\newcommand{\hth}{\hat\theta}
\newcommand{\rgt}{\rho_{\mathrm{gt}}}
\newcommand{\cgt}{c_{\mathrm{gt}}}
\newcommand{\Bp}[1]{B'_{\theta,#1}}

\title[Cooperative Learning for Distributed Online Optimization]{Cooperative Learning of Shared Latent Dynamics for Distributed Online Optimization}

\author{\Name{Shivanshu Tripathi}\Email{strip008@ucr.edu}\\
\addr Department of Electrical and Computer Engineering, UC Riverside, CA, USA\\
\Name{Maziar Raissi}\Email{maziarr@ucr.edu}\\
\addr Department of Mathematics, UC Riverside, CA, USA}

\begin{document}
\maketitle

\begin{abstract}
We study networked agents that must agree on a decision minimizing the sum of their local costs, where each cost contains unmeasurable stochastic parameters that evolve with linear but unknown dynamics shared by all agents. Each agent receives only finitely many noisy gradient measurements and must then keep tracking the moving minimizer. Our scheme reconstructs the parameters from windows of gradients, learns the shared dynamics cooperatively by averaging instrumental-variable moment matrices over the network, and tracks the minimizer by gradient tracking on forecast gradients. We bound the error of every agent in learning the dynamics and in tracking the minimizer, and show that the learning error, and through it the tracking error, decreases with the number of agents. We demonstrate the approach on two case studies. %
\end{abstract}


\section{Introduction}

Online optimization is a fundamental tool for real-time decision 
making, with applications ranging from the control of robots to 
content recommendation, in which the cost to be minimized 
changes over time and its minimizer must be tracked as the 
cost function changes \citep[e.g., see][]{AS-ED-SP-GL-GBG:20,
EDA-AS-SB-LM:120,PMW-MP-YH-AE-NM-ADP:24}. While existing methods
track a moving minimizer with guarantees when the model of the parameters is known or
gradients are available at all times, and learn unknown dynamics with guarantees when
the states are measured, they offer no guarantees when the parameters are latent, the data
are finite, and the information is dispersed and private, which is the typical situation in
networked systems. This poses critical limitations on the deployment of online optimization
for multi-agent decision making.

In this paper we propose a cooperative decision making scheme for 
multi-agent online optimization with the cost function having latent time-varying parameters. 
These agents cooperatively take a common decision. 
The latent parameters are observed only through finitely
many noisy gradient measurements, so their shared dynamics must be learned in order to
keep tracking the optimal decision once the measurements stop. Each agent reconstructs its
parameters from windows of measurements, and the agents learn the shared dynamics by
averaging instrumental-variable (IV) moment matrices over the network. We then
 derive bounds on the error of every agent in learning the dynamics and in tracking the
minimizer. Our bounds depend on the number
of agents, size of the training data, number of consensus rounds, and window
length. Finally, we illustrate our findings on two case studies.

\paragraph{Related work.}
Online learning in non-stationary environments tracks a moving minimizer
of a time-varying cost up to a constant error
\citep{OB-YG-AZ:15,AM-SS-AJ-AR:16,EH:16,SSS:12}. Time-varying optimization
improves on this with
prediction--correction \citep{AS-AM-AK-GL-AR:16,AS-ED:17,MF-SP-VMP-AR:17},
contracting dynamics \citep{AD-VC-AG-GR-FB:23}, or the internal model
principle \citep{GB-BVS:25a,GB-BVS:24,NB-RC-SZ:24,UC-NB-RC-SZ:25}, which
embeds a known model of the parameter dynamics. Our prior work
learned the model for a single agent \citep{ST-AAM-FP:24,ST-FP-MR:25}.
Unlike distributed subgradient methods \citep{AN-AO:09}, gradient tracking
\citep{JX-SZ-YCS-LX:15,PDL-GS:16} converges linearly with constant step sizes
\citep{AN-AO-WS:17,GQ-NL:18}, and online distributed methods track the moving minimizer
\citep{AS-AK-AM-GL-AR:17,SS-AJ:18,GC-FF-IN-GN:23,IN-AS-FF-GN:23}, but they need
the local gradients at every time. In contrast, our agents receive only
finitely many gradients. Non-asymptotic system identification
\citep{MS-HM-ST-MIJ-BR:18,SO-NO:19,TS-AR:19,AT-GP:19,IZ-ST:22,IZ-AT-BL-YJ-NM-GP:23}
and multi-task, federated, and meta-learning methods
\citep{HW-LFT-JA:23,LFT-HW-JA:23,AM-MKSF-AT-GM:24,LFT-DZ-JA-HW:24} show
that pooling data from similar systems reduces the sample complexity, but
assume measured states or outputs, whereas ours are observed through noisy gradients. In
contrast to federated learning \citep{HBM-EM-DR-SH-BAA:17}, which aggregates at a 
server and consensus-based fusion \citep{LX-SB-SL:05}, which estimates a static quantity,
our agents share only the messages with their neighbors, and the data remains private.

Throughout the paper, $\lVert \cdot \rVert$ is the Euclidean norm of a vector and the spectral 
norm of a matrix, and $\lVert \cdot \rVert_F$ is the Frobenius norm. $\sigma_{\text{min}}(\cdot),
\lambda_{\text{min}}(\cdot)$, and $\lambda_{\text{max}}(\cdot)$ are the smallest singular 
value, and the minimum and maximum eigenvalues. $\rho(\cdot)$ is the spectral radius, 
$(x)_+:=\max\{x,0\}$, and $\text{col} (v_1,\cdots,v_m)$ stacks the vectors $v_1,\cdots,v_m$.

\section{Problem setup}\label{sec:problem}

Consider $m$ agents that communicate over an undirected graph 
$\mc G=(\mc V,\mc E)$, with $\mc V=\{1,\dots,m\}$, and with neighborhoods $\mc N_i$, 
and cooperatively solve the time-varying problem: 
\begin{equation}\label{eq:prob}
\min_{x\in\R^n}~F(x,\theta(t)):=\sum_{i=1}^m f_i(x,\theta_i(t)),\qquad
f_i(x,\theta_i):=f_i^0(x)+g_i(x)^\T\theta_i,
\end{equation}
where $f_i^0:\R^n\to\R$ and $g_i:\R^n\to\R^p$ are known and twice
continuously differentiable, $\theta_i(t)\in\R^p$ is a latent parameter
private to agent $i$, and $\theta(t):=[\theta_1(t)^\T,\dots,\theta_m(t)^\T]^\T\in\mathbb{R}^{mp}$. The parameters evolve
under unknown linear stochastic dynamics common to all agents as:
\begin{equation}\label{eq:dyn}
\theta_i(t+1)=A\theta_i(t)+w_i(t),\qquad w_i(t)\sim\mc N(0,Q_i),\qquad i\in\mc V,
\end{equation}
with unknown $A\in\R^{p\times p}$ and $Q_i\succeq0$. Agent $i$ has access only
to a local gradient oracle:
\begin{align}\label{eq:grad}
y_i(x,t)=\nabla_xf_i^0(x)+C_i(x)\theta_i(t)+w_{m,i}(t)
\end{align}
with
$w_{m,i}(t)\sim\mc N(0,R_i)$, $R_i\succ0$, and $C_i(x):=\partial g_i^\T/\partial x\in\R^{n\times p}$
known; $\theta_i(0)$, $w_i(t)$, and $w_{m,i}(t)$ are mutually independent
across agents and time. Since $\nabla_xf_i^0(x)$ is known, agent $i$ can compute $\tilde y_i(x,t):=y_i(x,t)-\nabla_xf_i^0(x)=C_i(x)\theta_i(t)+w_{m,i}(t)$.

\paragraph{Data collection:}
We collect the data for $t\in\{0,\dots,N-1\}$, where agent $i$
queries its oracle at designed exploration points $x_i^{\mathrm e}(t)$ and
stores $\tilde y_i(t):=\tilde y_i(x_i^{\mathrm e}(t),t)$ and
$C_{i,t}:=C_i(x_i^{\mathrm e}(t))$. During the \emph{tracking phase} $t\ge N$
the oracle is no longer available.%

Agent $i$ knows only $f_i^0$, $g_i$, $R_i$ and its own data, and
exchanges messages with its neighbors. The goal is to bound the expected
tracking error $\E\norm{x_i(t)-x^*(t)}$ of every agent for $t\ge N$, where
$x^*(t)$ minimizes~\eqref{eq:prob}, and show that it decreases with $N$
and $m$, when the agents cooperatively estimate the dynamics $A$ of the system.
We make the following assumptions:

\begin{itemize}
\item[(A1)] There exist $0<\mu\le L$ and $\theta_i(t)\in\Theta_i\subseteq\R^p$ for all $t\geq 0$ and $i\in\mathcal{V}$,
such that $\nabla_x^2F(x,\theta)\succeq\mu I_n$ and $\norm{\nabla_x^2f_i(x,\theta_i)}\le L$
for all $x\in\R^n$, $\theta\in\Theta$, and $i\in\mc V$. 
\item[(A2)] $A$ is Schur stable, i.e., spectral radius $\rho(A)<1$, and invertible, and $(A,Q_i^{1/2})$ is controllable for all $i\in\mathcal{V}$. %
\item[(A3)] The exploration points lie in a bounded set, and there exist
$k_0\ge\lceil p/n\rceil$ and $\alpha_0>0$ with
$\sum_{j=0}^{k_0-1}C_{i,t+j}^\T C_{i,t+j}\succeq\alpha_0I_p$ for all $i\in\mc V$
and $t\in\{0,\dots,N-k_0\}$.
\item[(A4)] $\mc G$ is connected, and $W\in\R^{m\times m}$ is symmetric and doubly
stochastic with $W_{ij}>0$ if and only if $j\in\mc N_i\cup\{i\}$.
\end{itemize}

Assumption~(A1) makes the cost function $F(\cdot,\theta)$ as $\mu$-strongly convex for
every $\theta\in\Theta$, with a unique minimizer
$x^*(\theta)$. Since $\nabla_xF(x^*(\theta),\theta')=C(x^*(\theta))(\theta'-\theta)$
with $C(x):=[C_1(x)\cdots C_m(x)]$, for
$\theta,\theta'\in\Theta$ we have: 
\begin{equation}\label{eq:lip}
\norm{x^*(\theta')-x^*(\theta)}\le L_*\norm{\theta'-\theta},\quad
\text{where~~~~~~}L_*:=\frac{1}{\mu}\textstyle\sup_{\theta\in\Theta}\norm{C(x^*(\theta))}.
\end{equation}
Assumption~(A2) guarantees that the dynamics \eqref{eq:dyn} have a unique 
stationary covariance $\Sigma_{\theta,i}\succ 0$ satisfying $\Sigma_{\theta,i}=A\Sigma_{\theta,i} A^\T+Q_i$, and that there exists $C_\rho\geq 1$ and $\rho_A\in(\rho(A),1)$ with $\lVert A^j\rVert\leq C_\rho \rho_A^j$ for all $j\geq 0$. We define
$P_i(t):=\E[\theta_i(t)\theta_i(t)^\T]\succeq\Sigma_{\theta,i}$. 
Assumption~(A3)
is a persistence-of-excitation condition on the exploration sequence
with known state-dependent matrix $C_i(x)$. By Assumption~(A4),
$\sigma:=\norm{W-J}<1$, where $J:=\ones_m\ones_m^\T/m$.

\section{Cooperative identification and tracking}\label{sec:alg}

The proposed approach has three steps. First, a local reconstruction step,
which computes the Gauss--Markov window estimator of the latent
time-varying parameter $\theta_i(t)$. Second, a cooperative identification
step, which averages the instrumental-variable moments of all
agents over the network. Third, a network optimization step,
which is a gradient-tracking iteration driven by the oracle
during data collection and by the local forecasts afterwards. 
The details of each step are presented next. We summarize the 
overall approach in Algorithm~\ref{alg:main}. 

\paragraph{Step 1\!\!}\!\!{(Local reconstruction of $\theta_i(t)$)}:
For $t<N$, for a window length $k\ge k_0$, agent $i$ stacks the measurements
of each window $\{t,\ldots,t+k-1\}$, $\bar y_{i,t}:=[\tilde y_i(t)^\T,\dots,\tilde y_i(t+k-1)^\T]^\T$,
$\bar C_{i,t}:=[C_{i,t}^\T,\dots,C_{i,t+k-1}^\T]^\T$, $\bar R_i:=I_k\otimes R_i$,
and computes the window estimates of \citet{ST-FP-MR:25} as
\begin{equation}\label{eq:win}
\tth_i(t):=K_{i,t}\bar y_{i,t},\qquad
K_{i,t}:=(\bar C_{i,t}^\T\bar R_i^{-1}\bar C_{i,t})^{-1}\bar C_{i,t}^\T\bar R_i^{-1},
\qquad t\in\{0,\dots,N-k\},
\end{equation}
where the Gauss--Markov estimator neglects the drift of
$\theta_i$ inside the window. Since
$\theta_i(t+j)=\theta_i(t)+(A^j-I)\theta_i(t)+\sum_{l<j}A^{j-1-l}w_i(t+l)$
and $K_{i,t}\bar C_{i,t}=I_p$, we have
\begin{equation}\label{eq:decomp}
\tth_i(t)=\theta_i(t)+\eta_i(t)+\beta_i(t),
\end{equation}
with the noise $\eta_i(t):=K_{i,t}\bar w_{m,i,t}$, where $\bar w_{m,i,t}$
stacks $w_{m,i}(t),\dots,w_{m,i}(t+k-1)$, and the bias
$\beta_i(t):=K_{i,t}\bar b_{i,t}$, where $\bar b_{i,t}$ stacks
$C_{i,t+j}(\theta_i(t+j)-\theta_i(t))$, $j=0,\dots,k-1$. The noise is zero-mean
Gaussian and independent of $\eta_i(s)$ for $|t-s|\ge k$, the bias is zero
mean and independent of the noise, and both are independent across agents.

\paragraph{Step 2\!\!}\!\!{(Cooperative identification):}
Substituting~\eqref{eq:decomp} into~\eqref{eq:dyn} gives, for every agent:
\begin{equation}\label{eq:reg}
\tth_i(t+1)=A\tth_i(t)+\xi_i(t),\qquad
\xi_i(t):=\underbrace{w_i(t)}_{\text{process noise}}\!\!\!\!\!+\underbrace{\eta_i(t+1)-A\eta_i(t)}_{\text{noise residual}}+\underbrace{\beta_i(t+1)-A\beta_i(t)}_{\text{bias residual}}.
\end{equation}
Least squares on~\eqref{eq:reg} is inconsistent, since the regressor and the residual
share $\eta_i(t)$, so that $\E[\xi_i(t)\tth_i(t)^\T]=-A\Sigma_{\eta,i}(t)\ne0$. Using the 
instrument $z_i(t):=\tth_i(t-k)$, the local moments are: 
\begin{equation}\label{eq:moments}
\hat M_{j,i}:=\tfrac1{N-2k}\textstyle\sum_{t=k}^{N-k-1} \tth_i(t+j)z_i(t)^\T,\qquad j\in\{0,1\}.
\end{equation}
Let $M_{j,i}:=\E[\hat M_{j,i}]$. Since $w_i(t)$ and $\eta_i(t+1)-A\eta_i(t)$
are independent of $z_i(t)$, $M_{1,i}=AM_{0,i}+\Xi_i$ with
$\Xi_i:=\frac{1}{N-2k}\sum_{t}\E[(\beta_i(t+1)-A\beta_i(t))z_i(t)^\T]$, for every
$i$ with the \emph{same} $A$; hence the pooled moments
$\bar M_j:=\frac1m\sum_iM_{j,i}$ satisfy $\bar M_1=A\bar M_0+\bar\Xi$, with $\bar\Xi:=\frac{1}{m}\sum_i \Xi_i$ and
$\bar M_1\bar M_0^{-1}$ identifies $A$ up to the bias residual
$\bar\Xi\bar M_0^{-1}$. The agents approximate the pooled sample moments by
$\tau$ rounds of average consensus and estimate the dynamics as:
\begin{equation}\label{eq:cons}
\hat M_j^{(i)}(r+1)=\textstyle\sum_{l\in\mc N_i\cup\{i\}}W_{il}\hat M_j^{(l)}(r),\qquad
\Ai:=\hat M_1^{(i)}(\tau)\big(\hat M_0^{(i)}(\tau)\big)^{-1},
\end{equation}
for $r=0,\dots,\tau-1$, with $\hat M_j^{(i)}(0)=\hat M_{j,i}$.

\paragraph{Step 3\!\!}\!\!{(Network optimization):}
For $t\ge N$, agent $i$ propagates its last window estimate through the
identified dynamics
and forms the forecast gradient
\begin{equation}\label{eq:fore}
\hth_i(t):=(\Ai)^{t-(N-k)}\tth_i(N-k),\qquad
\hat y_i(x,t):=\nabla_xf_i^0(x)+C_i(x)\hth_i(t).
\end{equation}
The predicted minimizer is
$\hat x^*(t)$. 
Every agent maintains a
decision $x_i(t)$ and a gradient tracker $s_i(t)$ updated by the
gradient-tracking recursion \citep{AN-AO-WS:17} in its adapt-then-combine
form \citep{JX-SZ-YCS-LX:15}, with sums
over $j\in\mc N_i\cup\{i\}$ as
\begin{equation}\label{eq:gt}
x_i(t+1)=\textstyle\sum_{j}W_{ij}\big(x_j(t)-\alpha s_j(t)\big),\qquad
s_i(t+1)=\textstyle\sum_{j}W_{ij}\big(s_j(t)+\gamma_j(t+1)-\gamma_j(t)\big),
\end{equation}
with step size $\alpha>0$, $s_i(0)=\gamma_i(0)$, and the gradient signal $\gamma_i(t)$
equal to the oracle gradient $y_i(x_i(t),t)$ for $t<N$ and to the forecast gradient
$\hat y_i(x_i(t),t)$ for $t\ge N$. 

\begin{algorithm2e}[t]
\footnotesize
\caption{Cooperative identification and tracking}\label{alg:main}
\KwIn{$f_i^0$, $g_i$, $R_i$, $\{x_i^{\mathrm e}(t)\}_{t<N}$, $k$, $\tau$, $\alpha$, $x_i(0)$, $\gamma_i(0)=y_i(x_i(0),0)$, $s_i(0)=\gamma_i(0)$}
\For{$t=0,1,2,\dots$}{
  \lIf{$t<N$}{query the oracle at $x_i^{\mathrm e}(t)$; store $\tilde y_i(t)$}
  exchange $x_i(t)-\alpha s_i(t)$ with $\mc N_i$ and update $x_i(t+1)$ by~\eqref{eq:gt}\;
  \lIf{$t+1<N$}{$\gamma_i(t+1)=y_i(x_i(t+1),t+1)$ \tcp*[f]{oracle gradient}}
  \If{$t+1=N$}{compute $\{\tth_i(t')\}_{t'\le N-k}$ by~\eqref{eq:win} and $\hat M_{0,i},\hat M_{1,i}$ by~\eqref{eq:moments}; 

  run $\tau$ consensus rounds and set $\Ai$ by~\eqref{eq:cons}}
  \lIf{$t+1\ge N$}{forecast $\hth_i(t+1)$ by~\eqref{eq:fore}; $\gamma_i(t+1)=\hat y_i(x_i(t+1),t+1)$ \tcp*[f]{forecast gradient}}
  exchange $s_i(t)+\gamma_i(t+1)-\gamma_i(t)$ with $\mc N_i$ and update $s_i(t+1)$ by~\eqref{eq:gt}\;
}
\end{algorithm2e}

\section{Main results}\label{sec:theory}

For $t\ge N$ the error of agent $i$ splits as
\begin{equation}\label{eq:split}
\norm{x_i(t)-x^*(t)}\le\underbrace{\norm{x_i(t)-\bar x(t)}}_{\text{disagreement}}
+\underbrace{\norm{\bar x(t)-\hat x^*(t)}}_{\text{optimization}}
+\underbrace{\norm{\hat x^*(t)-x^*(t)}}_{\text{prediction}},
\end{equation}
where $\bar x(t):=\frac1m\sum_ix_i(t)$. By~\eqref{eq:lip} 
$$\norm{\hat x^*(t)-x^*(t)}\leq L_*\sum_i\norm{\hth_i(t)-\theta_i(t)},$$ and, with
$H:=t-(N-k)$, $\tilde e_i:=\tth_i(N-k)-\theta_i(N-k)$, and
$\theta_i(t)=A^H\theta_i(N-k)+\sum_{j<H}A^jw_i(t-1-j)$, the forecast error of
agent $i$ decomposes as
\begin{equation}\label{eq:fdec}
\hth_i(t)-\theta_i(t)=\underbrace{\big((\Ai)^H-A^H\big)\tth_i(N-k)}_{\text{identification}}
+\underbrace{A^H\tilde e_i}_{\text{reconstruction}}
-\underbrace{\textstyle\sum_{j<H}A^jw_i(t-1-j)}_{\text{process noise}}.
\end{equation}
We bound each error in the subsequent subsections. All proofs of lemmas 
and theorems are given in the supplementary material.

\subsection{Reconstruction and cooperative identification}\label{sec:th-id}
The estimate $\tilde \theta_i(t)$ in \eqref{eq:win} has a 
reconstruction error $\mathbb{E}\lVert \tilde\theta_i(t)-\theta_i(t) \rVert_2$. 

\begin{lemma}[Reconstruction error] \label{lemma: reconstruction}
Let (A2)--(A3) hold, $\Gamma_i:=\sup_{x\in \mathbb{R}^n} \lVert C_i(x) \rVert_2$, $B_{\theta,i}:=
\sqrt{\trace(\Sigma_{\theta,i})}$, and $k\geq k_0$.  For all
$i\in\mc V$ and $t\in\{0,\dots,N-k\}$:
\begin{enumerate}
\item $\bar C_{i,t}^\T\bar R_i^{-1}\bar C_{i,t}\succeq\alpha_{k,i}I_p$ with
$\alpha_{k,i}:=\lfloor k/k_0\rfloor\alpha_0/\lmax(R_i)$.
\item The covariance
$\Sigma_{\eta,i}(t)$ of $\eta_i(t)$ satisfies $\Sigma_{\eta,i}(t)\preceq\alpha_{k,i}^{-1}I_p$.
\item For $\chi_i^2:=2k_0\lmax(R_i)/(\alpha_0\lmin(R_i))$, the reconstruction error satisfies 
\begin{equation}\label{eq:recon}
\big(\E\norm{\tth_i(t)-\theta_i(t)}^2\big)^{1/2}\le\sqrt{p/\alpha_{k,i}}+\Gamma_iC_\rho\chi_i\Big(\tfrac{k}{\sqrt3}\norm{A-I}B_{\theta,i}+\sqrt{\tfrac k2\trace(Q_i)}\Big):=\ew_i.
\end{equation}
\end{enumerate}
\end{lemma}

 We identify the common dynamics $A$ using the sample moments \eqref{eq:moments}, 
 which every agent computes locally and then averages.  The moment of agent $i$ is given as $ M_{0,i} = A^k \Sigma_{\theta,i}+ \Delta_{0,i}$, where $A^k \Sigma_{\theta,i}=\mathbb{E}[\theta_i(t)\theta_i(t-k)^\T]$ is the correlation between the regressor $\tilde \theta_i(t)$ and the lagged instrument 
 $\tilde \theta_i(t-k)$, and $\Delta_{0,i}$ collects the terms involving the windowing bias $\beta_i$.
 The pooled moment is $\bar M_{0} = A^k \bar\Sigma_{\theta}+ \bar\Delta_{0}$ with
 $\bar \Sigma_{\theta} := \frac{1}{m}\sum_i\Sigma_{\theta,i}$ and $\bar \Delta_0=\frac{1}{m}\sum_i \Delta_{0,i}$. We define the
 invertibility margin $\nu_k:=\sigma_{\text{min}} (A^k \bar\Sigma_\theta)>0$ of $\bar M_0$, 
 and the heterogeneity $D_M:=\max_{j\in \{0,1\},l\in\mathcal{V}} \lVert M_{j,l} - \bar M_j \rVert_2$, 
 which vanishes for identical agents. 

 The estimate $\hat A^{(i)}$ differs from $A$ for three reasons. 
 First, the moments are computed from finitely many samples (finite sample error). Second, after 
 $\tau$ consensus rounds, agent $i$ knows the network average only up to a consensus error of order $\sigma^\tau$. Third, the  
error due to bias residual $\bar\Xi \bar M_0^{-1}$ contributes an error that does not vanish with $N$. 

\begin{theorem}[Identification error]\label{thm:id}
Let Assumptions~(A2)--(A4) hold, $k\ge k_0$,
$\delta\in(0,1/2]$, $N\ge N_0(\delta/(4m))$, $\norm{\bar\Delta_0}\le\nu_k/2$,
and $\epsm(N,\tau,\delta)\le\nu_k/8$, where
\begin{equation}\label{eq:em}
\epsm(N,\tau,\delta):=c\sqrt{\log(4m/\delta)/N}\,\big(m^{-1/2}+\sigma^\tau\big)+\sqrt m\,\sigma^\tau D_M.
\end{equation}
where $c>0$, independent of $N$, $m$ and $\delta$, is the \emph{concentration constant} of the
sample moments \citep{MR-RV:13}: for all $j\in\{0,1\}$, $\omega\in\mathbb R^m$ and $\delta\in(0,1)$,
\begin{align}\label{eq:conc}
\Big\lVert\textstyle\sum_{l=1}^m\omega_l\big(\hat M_{j,l}-M_{j,l}\big)\Big\rVert\leq c\,\lVert\omega\rVert\sqrt{\tfrac{\log(1/\delta)}{N}}.
\end{align}
Then there is an event $\mc E_\delta$ with $\Prob(\mc E_\delta)\ge1-\delta$ on
which, for all $i\in\mc V$ simultaneously,
\begin{align}\label{eq:iv}
\begin{aligned}
\norm{\Ai-A}\le\underbrace{C_A(1+\sqrt{m}\sigma^\tau) \sqrt{\frac{\log(4m/\delta)}{mN}}}_{\text{finite-sample error}}  + \underbrace{C_W\sqrt{m} \sigma^\tau D_M}_{\text{heterogeneity}} + &\underbrace{C_B(k\lVert A-I \rVert_2 + \sqrt{k\trace(\bar Q)})}_{\text{bias residual}}\nonumber\\&=:\epsilon^{\text{id}}(\delta),
\end{aligned}
\end{align}
 where $\bar Q:=\frac{1}{m} \sum_i Q_i$, and the constants $C_A,C_B,C_W$ are independent of $N,\tau$ and $\delta$.
\end{theorem}

\subsection{Tracking}\label{sec:th-track}
For $t\ge N$ the oracle is no longer available and agent $i$ replaces its
gradient by the forecast gradient $\hat y_i(x,t)$ of~\eqref{eq:fore}, which
is the gradient of the \emph{forecast cost}
$\hat f_i(x,t):=f_i^0(x)+g_i(x)^\T\hth_i(t)$. The algorithm tracks
the minimizer $\hat x^*(t)$ of $\sum_i\hat f_i(\cdot,t)$. 
Let $x(t):=\text{col}(x_1(t),\ldots,x_m(t))$, $s(t):=\text{col}(s_1(t),\ldots,s_m(t))$, $\hat\theta(t):=\text{col}(\hat\theta_1(t),\ldots,\hat\theta_m(t))$, and $\bar s(t):=\frac{1}{m}\sum_i s_i(t)$. We define:
\begin{align}
\begin{aligned}
a_t&:=\lVert x(t)-\ones_m\otimes\bar x(t)\rVert_2&&\text{(iterate disagreement)},\\
b_t&:=\sqrt m\,\lVert\bar x(t)-\hat x^*(t)\rVert_2&&\text{(optimization error)},\\
c_t&:=\lVert s(t)-\ones_m\otimes\bar s(t)\rVert_2&&\text{(tracker disagreement)},\\
d_t&:=\lVert\hat\theta(t+1)-\hat\theta(t)\rVert_2&&\text{(forecast drift)},
\end{aligned}
\end{align}
with $e_t:=[a_t\ \ b_t\ \ c_t]^\T$. 
The first two are the disagreement and
optimization terms of~\eqref{eq:split}, $c_t$ measures how well the trackers
follow the average gradient, and $d_t$ bounds the motion of the predicted
minimizer, since $\norm{\hat x^*(t+1)-\hat x^*(t)}\le L_*d_t$
by~\eqref{eq:lip}. 
\begin{lemma}[Gradient tracking]\label{lemma:3}
Let (A1)--(A4) hold, let $\kappa:=mL/\mu$, $\Gamma:=\max_{i\in\mathcal{V}}\Gamma_i$, and let
$\alpha\in(0,\alpha_{\max}]$, where $\alpha_{\max}:=\min\Big\{\frac{(1-\sigma)^2}{4L},\ \frac{1}{(\kappa+1)L}\Big\}$.
Then there exist $\rho_{\rm gt}\in(0,1)$ and $c_{\rm gt}>0$, %
such that for all $i\in\mathcal V$ and $t\geq N$,
\begin{align}\label{eq:gtbound}
\lVert x_i(t)-\hat x^*(t)\rVert_2\leq c_{\rm gt}\Big(\rho_{\rm gt}^{\,t-N}\lVert e_N\rVert_2+\!\!\sum_{q=N}^{t-1}\!\rho_{\rm gt}^{\,t-1-q}d_q\Big).
\end{align}
\end{lemma}
The iterates forget their initial state at the rate $\rgt$ and lag behind
the predicted minimizer by a weighted sum of past drifts. The
drift is produced by forecasts as
 next theorem shows.

The tracking error of agent $i$ thus has three sources. First, the network
needs time to reach consensus and to optimize. Second, the predicted minimizer moves while the
forecasts are propagated, so the iterates lag behind it. Third, the
forecasts are wrong, by the identification, reconstruction, and
process-noise terms of~\eqref{eq:fdec}, which shifts the minimizer
away from $x^*(t)$. Since the identification results
hold on an event $\mc E_\delta$. For a random variable $Z$, we have
$\Ed[Z]:=\E[Z\ind_{\mc E_\delta}]$ for $Z\ge0$, so that
$\E[Z\mid\mc E_\delta]\le\Ed[Z]/(1-\delta)$. 

\begin{theorem}[Distributed tracking error]\label{thm:main}
Let Assumptions~(A1)--(A4) and the hypotheses of
Theorem~\ref{thm:id} and Lemma~\ref{lemma:3} hold, and let
$\hat\rho:=\rho_A+C_\rho\eid(\delta)<1$ and $\rho_\star:=\max\{\rgt,\hat\rho\}$.
Then, for every agent $i\in\mc V$ and every $t\ge N$,
\begin{equation}\label{eq:main}
\begin{aligned}
\Ed\norm{x_i(t)-x^*(t)}\le{}&
\underbrace{\cgt\rgt^{\,t-N}\Ed\norm{e_N}}_{\text{(i) optimization error}}
+\underbrace{\cgt C_\rho\hat\rho^{\,k}(t-N)\rho_\star^{\,t-N-1}\,U_N}_{\text{(ii) forecast drift}}+\underbrace{L_*\textstyle\sum_{j=1}^m\ef_j(N,t;\delta)}_{\text{(iii) forecast error}},
\end{aligned}
\end{equation}
where $U_N:=\big(\norm{A-I}+\eid(\delta)\big)\sum_{j=1}^m\big(B_{\theta,j}+\ew_j\big)$,
$\Ed\norm{e_N}<\infty$, and $\ef_j(N,t;\delta)$ bounds the forecast error of
agent $j$ with $H=t-(N-k)$ and
$\sigma_{H,j}^2:=\trace\sum_{l<H}A^lQ_j(A^l)^\T\le\trace\Sigma_{\theta,j}$,
\begin{equation}\label{eq:fore-bound}
\begin{aligned}
\underbrace{\Ed\norm{\hth_j(t)-\theta_j(t)}}_{\text{forecast error}} \le 
\underbrace{C_\rho^2H\hat\rho^{\,H-1}\big(B_{\theta,j}+\ew_j\big)\,\eid(\delta)}_{\text{identification}}+\underbrace{C_\rho\rho_A^{\,H}\,\ew_j}_{\text{reconstruction}}
+\underbrace{\sigma_{H,j}}_{\text{process noise}}\!\!\!\!\!.
\end{aligned}
\end{equation}
\end{theorem}

\section{Numerical experiments}\label{sec:exp}

We evaluate the proposed methodology on two case studies: (i) charging-cart placement for patrolling UAVs, simulated in Webots, and (ii) decentralized recommendation with regional servers. The project page and source code are publicly available at \href{https://shivanshutripath.github.io/cooperative-learning/}{video and code}.

\subsection{A charging cart for patrolling UAVs}\label{sec:5.1}

\begin{figure}[t]
  \centering
  \includegraphics[width=\linewidth]{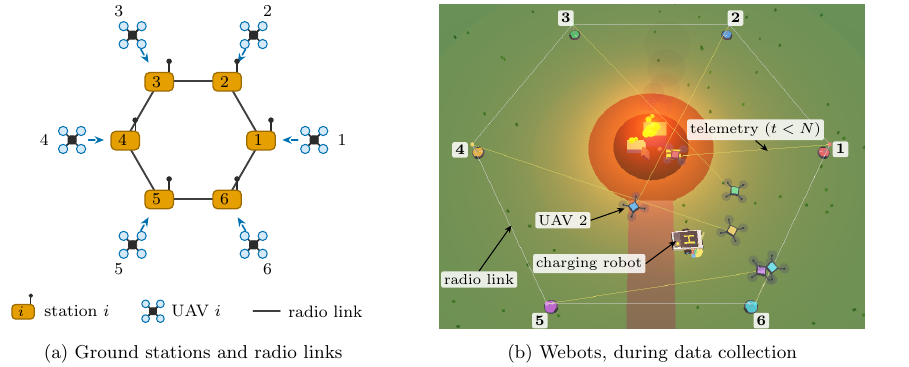}
\caption{The setting of Section~\ref{sec:5.1}. (a) Station~$i$
  receives the telemetry of UAV~$i$ only, while $t<N$, and talks only to its
  two neighbors on the ring. (b) The Webots simulation setup with each UAV 
  having the color of its station.}
  \label{fig:rv-setup}
\end{figure}

Figure~\ref{fig:rv-setup} illustrates the setting and the Webots simulation setup. 
During the response to a wildfire, six
agencies (police, fire, medical, search and
rescue, utilities, and civil protection) each fly one patrol UAV over the
site. UAV~$i$ follows a fixed patrol route $p_i(t)$, known only to its
agency, whose ground station receives the
telemetry of that UAV only. The agencies share neither telemetry, nor video, nor routes{, and
each station exchanges messages only with its two neighbors on the ring of
Figure~\ref{fig:rv-setup}(a)}. A {generator-powered} ground robot {(the cart)} carries a charging {pad and, as is
usual for mobile recharging stations
need not be refueled during the mission}; a UAV
that needs power flies to the cart, charges, and returns to
its route. The stations must agree on where the cart should be as the UAVs move.
The position of UAV~$i$ is its planned position plus a deviation,
\begin{equation*}
b_i(t)=p_i(t)+\theta_i(t),\qquad \theta_i(t+1)=A\,\theta_i(t)+w_i(t).
\end{equation*}
The deviation $\theta_i(t)$ is unknown and wind
gusts $w_i(t)$ push the UAV off its route. {The UAVs are of the same type and fly in the
same wind field, so their} deviations share the dynamics
$A=\begin{bmatrix}0.979&-0.148\\0.148&\hphantom{-}0.979\end{bmatrix}$,
unknown to the stations.
{Placing} the cart at $x$ costs UAV~$i$ the squared distance to it, weighted
by {the known rate $h_i$} at which that UAV needs power{:}
\begin{equation*}
f_i(x,\theta_i)=h_i\norm{x-b_i(t)}^2=h_i\norm{x-p_i(t)-\theta_i}^2 .
\end{equation*}
The team cost is smallest at the weighted centroid
$x^{{*}}(t)=\sum_ih_ib_i(t)/\sum_ih_i$, which moves with the UAVs.
{Up to a term independent of $x$, $f_i$ is as in~\eqref{eq:prob} with}
$f_i^0(x)=h_i\norm{x-p_i(t)}^2$ and $g_i(x)=-2h_i(x-p_i(t))$ {(time-varying only
through the known route)}, so
$C_i=-2h_iI_2$ is invertible and Assumption~(A3) holds with
$k_0=1$. A noisy gradient of $f_i$ at any $x$ is a noisy position of
UAV~$i$, which station~$i$ receives once per second during data collection
($t<N$). At time $t=N$ {the telemetry is lost (e.g., in a cellular outage), and
each station must forecast the position of its UAV} to keep the cart in place.
The UAVs start on average $50$\,m off their routes, gusts displace them by
$0.5$\,m per second, and the received positions carry errors of $5$ to $15$\,m.

\begin{figure}[t]
  \centering
  \includegraphics[width=\linewidth]{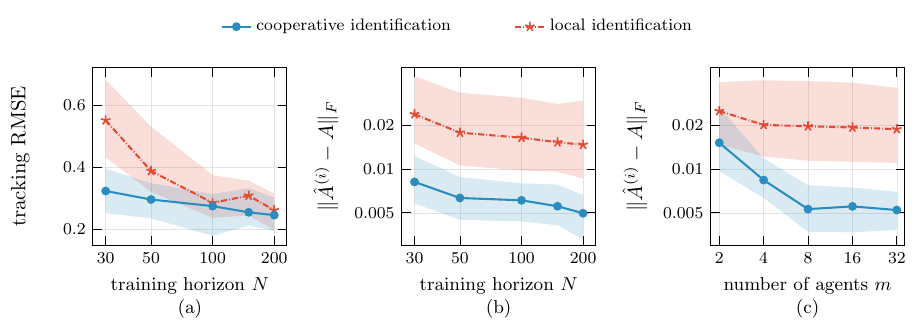}
\caption{Cooperative identification (solid blue) and local identification
  (dash-dotted red) for the setting of Section~\ref{sec:5.1}, over
  $50$ Monte Carlo trials with a tracking horizon of $200$ steps. (a) Tracking RMSE
  vs the training horizon $N$, (b) identification error $\|\hat A^{(i)}-A\|_F$
  vs the training horizon $N$, and (c) identification error versus the
  number $m$ of identical agents ($h_i=1$) and $N=100$.}
  \label{fig:rv}
\end{figure}

The iterates start at $x_i(0)=0$.
Cooperative identification {uses} $\tau=3$ consensus rounds and {local identification uses $\tau=0$}. 
Fig.~\ref{fig:rv}(a) shows {that} the tracking RMSE $\big(\frac{1}{150m}\sum_{t=N+50}^{N+199} \sum_{i=1}^m \lVert x_i(t)-x^*(t) \rVert^2\big)^{1/2}$ {decreases with $N$ and is lower with cooperation, by up to $41\%$ for $N=30$}. 
Fig.~\ref{fig:rv}(b) shows that cooperative identification is $2.7$--$3$
times more accurate than local identification for every $N$. 
The cooperative error decreases from $0.008$ to $0.005$ as $N$ grows. 
Fig.~\ref{fig:rv}(c) shows that the {cooperative error} decreases {with the number of agents} (as $1/\sqrt{m}${, by} Theorem~\ref{thm:id}) but saturates for $16$ or more agents, where $\sigma$ reaches $0.85$ and $0.96$ and $\sigma^\tau$ is no longer small{, while the error of local identification does not improve}.

\subsection{Decentralized recommendation with regional servers}\label{sec:exp-rc}

\begin{figure}[t]
  \centering
  \includegraphics[width=\linewidth]{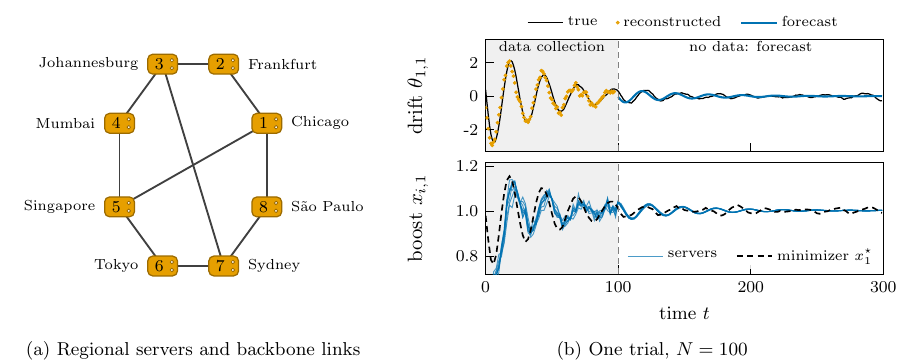}
\caption{The setting of Section~\ref{sec:exp-rc}. (a) {Server~$i$ receives} only the engagement data
  of region~$i${,} for $t<N$, and exchanges messages with its
  backbone neighbors. (b) {Top:} the drift
  $\theta_{1,1}$ for region~$1$ and genre~$1$, its reconstruction by server~$1$
  ($t<N$), and its forecast with the cooperative estimate of $A$
  ($t\ge N$). {Bottom:} the boosts $x_{i,1}$ of genre~$1$ at the eight servers
  and the minimizer $x^\star_1$ of the team cost.}
  \label{fig:rc-setup}
\end{figure}

Figure~\ref{fig:rc-setup} illustrates the setting with servers, users and links. {An} online marketplace
serves its users from $m=8$ regional data centers. Each regional server collects the data of the 
region, and there is no central aggregator, as in decentralized
federated learning \citep{AK-NL-SB-MJ-SUS:20}. Each server exchanges messages
only with {its neighbors in the backbone network of Figure~\ref{fig:rc-setup}(a)}. Only the
ranking model is shared, so the servers must agree on common boosts
$x\in\R^3${, one for each of the top three product genres (e.g., outdoor, home, and tech)}.
The interest of the users of region~$i$ in genre~$\ell$ is a known average
plus a drift,
\begin{equation*}
u_{i,\ell}(t)=\bar u_{i,\ell}+\theta_{i,\ell}(t),\qquad
\theta_i(t+1)=A\,\theta_i(t)+w_i(t).
\end{equation*}
The drift $\theta_i(t)$ is unknown and private to server~$i$, and the noise 
$w_i(t)$ models short-term fluctuations of taste. The interests of all
regions follow the same trend dynamics
$$A=\begin{bmatrix}
0.9495 & -0.2425 & 0\\
0.2425 & 0.9495 & 0\\
0 & 0 & 0.95
\end{bmatrix},$$ 
which is unknown to the
servers. The dynamics shows a damped oscillation shared by the first two genres and a
fading interest in the third.
Under a boost $x_\ell$, a user engages with an item of genre $\ell$ with 
probability $P(x_\ell)$, where $P(z)=1/(1+e^{-z})$ is the logistic 
function, so the negative log-likelihood of the engagement is $-\log P(x_\ell)=\mathrm{softplus}(-x_\ell)$.
{Server~$i$ weights this loss by the interest of its users
and adds a ridge penalty} with weight $\lambda_i>0$, which keeps the boosts small:
\begin{equation*}
f_i(x,\theta_i)=\frac{\lambda_i}2\norm x^2
+\sum_{\ell=1}^3 u_{i,\ell}(t)\,\mathrm{softplus}(-x_\ell){.}
\end{equation*}
We consider
$f_i^0(x)=\frac{\lambda_i}2\norm x^2+\sum_{\ell}\bar u_{i,\ell}\,\mathrm{softplus}(-x_\ell)$
and $g_i(x)=\mathrm{softplus}(-x)$ (elementwise), so
$C_i(x)=-\diag(P(-x))$ is invertible for every $x$ and
Assumption~(A3) holds with $k_0=1$. 
The noise $w_i(t)$ has
standard deviation $0.02$, and the gradients carry noise of standard
deviation $0.15$. {Gradients are logged only for $t<N$; then the servers rely on forecasts.} %

\begin{figure}[t]
  \centering
  \includegraphics[width=\linewidth]{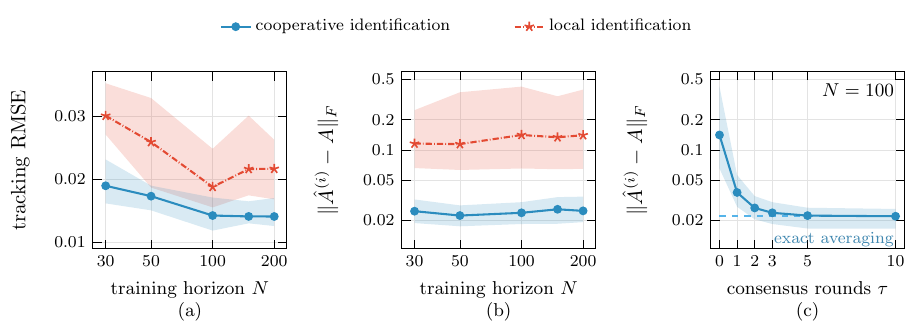} 
   \caption{Cooperative identification with $\tau=3$ (solid blue) and local identification $\tau=0$ (dash-dotted red)
for the setting of Section~\ref{sec:exp-rc}, over 50 Monte Carlo trials with a tracking horizon of 200 steps.
 (a) Tracking RMSE vs the
training horizon N, (b) identification error $\|\hat A^{(i)}-A\|_F$ versus $N$ and (c)
identification error versus the number of consensus rounds.}
  \label{fig:rc}
\end{figure}

We run $50$ Monte Carlo trials with the same tracking horizon,
initialization, simulation setting, and tracking RMSE as in Section~\ref{sec:5.1}. 
Figure~\ref{fig:rc-setup}(b) shows one trial with $N=100$. 
Fig.~\ref{fig:rc}(a) shows {that} the tracking RMSE with
cooperative identification is $24$--$37\%$ lower than with local
identification for every $N$. 
Fig.~\ref{fig:rc}(b) shows {that cooperation
reduces the identification error by a factor of} $4.7$--$5.9$. The cooperative
error stays between $0.022$ and $0.026$ for every $N$. 
Fig.~\ref{fig:rc}(c) varies the number of consensus rounds at $N=100$
($\sigma=0.68$). The identification error decreases from $0.141$ for
$\tau=0$ to $0.022$ for $\tau=5$. More details of the simulation along 
with the data used is given in the \href{https://shivanshutripath.github.io/cooperative-learning/}{project page}.

\paragraph{Summary.}
In both case studies, we observe that the cooperation reduces the identification error by the 
factor $\sqrt m$ of Theorem~\ref{thm:id}, or more when local identification
is unreliable. 
The cooperative error settles at the bias floor set by $k$
and $\norm{A-I}$ as shown in Theorem~\ref{thm:main}. 

\section{Conclusion}\label{sec:concl}
We studied online optimization over a network of agents whose local costs
depend on private latent parameters with shared unknown stochastic
dynamics, observed only through finitely many noisy gradients. Averaging
instrumental-variable moments over the network learns the dynamics{, and gradient tracking on
the forecast gradients tracks} the minimizer.
The simulations confirm the $\sqrt m$ gain in cooperative estimation, and show
that a few consensus rounds {suffice and} that cooperation
matters most when data are scarce. Future work includes considering directed 
or time-varying graphs, and considering non-linear dynamics.

\acks{This material is based upon work supported in part by award NSF-CMMI-2618914. }

\bibliography{l4dc2026-sample}

\clearpage
\appendix
\section{Supplementary material}\label{sec:supp}
\setcounter{equation}{0}
\renewcommand{\theequation}{S\arabic{equation}}
\renewcommand{\theHequation}{S\arabic{equation}}

\paragraph{Proof of Lemma~\ref{lemma: reconstruction}:}%

Fix $i\in\mc V$ and $t\in\{0,\dots,N-k\}$. Iterating~\eqref{eq:dyn}, for
$j=0,\dots,k-1$,
\begin{equation}\label{eq:drift}
\theta_i(t+j)=\theta_i(t)+\underbrace{(A^j-I)\theta_i(t)+\sum_{l=0}^{j-1}A^{j-1-l}w_i(t+l)}_{\Delta_{i,t}(j)},
\end{equation}
so the stacked measurements of the window satisfy:
\begin{equation}\label{eq:winmodel}
\begin{gathered}
\bar y_{i,t}=\bar C_{i,t}\theta_i(t)+\bar b_{i,t}+\bar w_{m,i,t},\\
\bar b_{i,t}:=\col\big(C_{i,t+j}\Delta_{i,t}(j)\big)_{j=0}^{k-1},\qquad
\bar w_{m,i,t}\sim\mc N(0,\bar R_i).
\end{gathered}
\end{equation}

\emph{(i) Excitation:} The window $\{t,\dots,t+k-1\}$ contains the
$\lfloor k/k_0\rfloor$ disjoint blocks $\{t+qk_0,\dots,t+(q+1)k_0-1\}$,
{$q=0,\dots,\lfloor k/k_0\rfloor-1$, each starting at a time
$t+qk_0\le N-k_0$, so that (A3) applies to each of them}. With
$\bar R_i^{-1}=I_k\otimes R_i^{-1}\succeq\lmax(R_i)^{-1}I_{kn}$, we have
\begin{equation}\label{eq:exc}
\bar C_{i,t}^\T\bar R_i^{-1}\bar C_{i,t}
\succeq\frac{1}{\lmax(R_i)}\sum_{q=0}^{\lfloor k/k_0\rfloor-1}\ \sum_{j=0}^{k_0-1}C_{i,t+qk_0+j}^\T C_{i,t+qk_0+j}
\succeq\frac{\lfloor k/k_0\rfloor\alpha_0}{\lmax(R_i)}I_p=\alpha_{k,i}I_p ,
\end{equation}
which is the first claim. 

\emph{(ii) Noise and bias:} 
$\eta_i(t)$ is a zero-mean Gaussian function of
$w_{m,i}(t),\dots,w_{m,i}(t+k-1)$ only, and, by~\eqref{eq:drift}, $\beta_i(t)$
is a zero-mean linear function of $\theta_i(t),w_i(t),\dots,w_i(t+k-2)$ only.
As these variables are independent, $\eta_i(t)$ is independent of every
$\beta_i(s)$, and of $\eta_i(s)$ whenever $|t-s|\ge k$; and the noises and
biases of different agents are independent. By~\eqref{eq:exc},
$\bar R_i\succeq\lmin(R_i)I_{kn}$, and $\lfloor k/k_0\rfloor\ge k/(2k_0)$,
\begin{equation}\label{eq:Seta}
\begin{gathered}
\Sigma_{\eta,i}(t)=K_{i,t}\bar R_iK_{i,t}^\T=\big(\bar C_{i,t}^\T\bar R_i^{-1}\bar C_{i,t}\big)^{-1}\preceq\alpha_{k,i}^{-1}I_p{,}\\
\norm{K_{i,t}}^2\le\frac{\norm{K_{i,t}\bar R_iK_{i,t}^\T}}{\lmin(R_i)}\le\frac{\lmax(R_i)}{\lfloor k/k_0\rfloor\,\alpha_0\,\lmin(R_i)}\le\frac{\chi_i^2}{k} .
\end{gathered}
\end{equation}
The first relation is the second claim.%

\emph{(iii) Bias:} $\theta_i(t)$ is independent of the later zero-mean
noises. Moreover,
$\norm{A^j-I}=\norm{(A-I)\sum_{l<j}A^l}\le C_\rho\,j\norm{A-I}$ and
$\E\norm{\sum_{l<j}A^{j-1-l}w_i(t+l)}^2=\sum_{l<j}\trace\big(A^{j-1-l}Q_i(A^{j-1-l})^\T\big)\le C_\rho^2\,j\trace Q_i$.
Hence
$\E\norm{\Delta_{i,t}(j)}^2\le C_\rho^2\big(j^2\norm{A-I}^2B_{\theta,i}^2+j\trace Q_i\big)$,
and with $\norm{C_{i,t+j}}\le\Gamma_i$, $\sum_{j<k}j^2\le k^3/3$, and
$\sum_{j<k}j\le k^2/2$,
\begin{equation}\label{eq:bbar}
\E\norm{\bar b_{i,t}}^2=\sum_{j=1}^{k-1}\E\norm{C_{i,t+j}\Delta_{i,t}(j)}^2
\le\Gamma_i^2C_\rho^2\Big(\frac{k^3}{3}\norm{A-I}^2B_{\theta,i}^2+\frac{k^2}{2}\trace Q_i\Big).
\end{equation}
By~\eqref{eq:Seta}, \eqref{eq:bbar}, and $\sqrt{a+b}\le\sqrt a+\sqrt b$,
\begin{equation}\label{eq:betabound}
\big(\E\norm{\beta_i(t)}^2\big)^{1/2}\le\norm{K_{i,t}}\big(\E\norm{\bar b_{i,t}}^2\big)^{1/2}
\le\frac{\chi_i}{\sqrt k}\,\Gamma_iC_\rho\Big(\frac{k^{3/2}}{\sqrt3}\norm{A-I}B_{\theta,i}+\frac{k}{\sqrt2}\sqrt{\trace Q_i}\Big)=b_{k,i}.
\end{equation}

\emph{(iv) Conclusion:} By~\eqref{eq:decomp} and Minkowski's inequality,
\begin{equation}\label{eq:Bprime}
\begin{gathered}
\big(\E\norm{\tth_i(t)-\theta_i(t)}^2\big)^{1/2}\le\big(\E\norm{\eta_i(t)}^2\big)^{1/2}+\big(\E\norm{\beta_i(t)}^2\big)^{1/2}\le\sqrt{p/\alpha_{k,i}}+b_{k,i}=\ew_i{,}\\
{\big(\E\norm{\tth_i(t)}^2\big)^{1/2}\le\big(\E\norm{\theta_i(t)}^2\big)^{1/2}+\big(\E\norm{\tth_i(t)-\theta_i(t)}^2\big)^{1/2}\le B_{\theta,i}+\ew_i=:\Bp{i}.}
\end{gathered}
\end{equation}
{The first bound is~\eqref{eq:recon} and the second is used in the proofs of
Theorems~\ref{thm:id} and~\ref{thm:main}.}

\paragraph{Proof of Theorem~\ref{thm:id}:}
The proof has five steps.

\emph{(i) Covariance of the window estimates:} Let
$\tilde\Theta_i:=\col(\tth_i(0),\dots,\tth_i(N-k))$. The exploration points,
and hence the gains $K_{i,t}$, are fixed in advance, so $\tilde\Theta_i$ is a
linear function of the measurements $\tilde y_i(0),\dots,\tilde y_i(N-1)$. 
Since the measurements are zero-mean
Gaussian, so is $\tilde\Theta_i$, and $\norm{\E[\tilde\Theta_i\tilde\Theta_i^\T]}$
is at most $\chi_i^2$ times the norm of the covariance of the measurements
{(the linear map from the measurements to $\tilde\Theta_i$ has norm at
most $\chi_i$, since each measurement enters at most $k$ of the windows and
$\norm{K_{i,t}}^2\le\chi_i^2/k$ by~\eqref{eq:Seta})}.
That covariance is the sum of a noise part, of norm $\lmax(R_i)$, and a
parameter part, of norm at most $\Gamma_i^2$ times that of the covariance of
$\col(\theta_i(0),\dots,\theta_i(N-1))$. The blocks
$\E[\theta_i(t)\theta_i(s)^\T]=A^{t-s}P_i(s)$, $t\ge s$, have norm
{at most} $C_\rho B_{\theta,i}^2\rho_A^{\,t-s}$, and a symmetric block matrix has
norm {at most the largest sum of the block norms along a block row, here
at most} $2C_\rho B_{\theta,i}^2/(1-\rho_A)$. Therefore
\begin{equation}\label{eq:cov}
\norm{\E[\tilde\Theta_i\tilde\Theta_i^\T]}\le c_1:=\max_{l\in\mc V}\chi_l^2\Big(\frac{2C_\rho\Gamma_l^2B_{\theta,l}^2}{1-\rho_A}+\lmax(R_l)\Big).
\end{equation}

\emph{(ii) Concentration of the moments:} An entry $(a,b)$ in~\eqref{eq:moments} is
$(\hat M_{j,i})_{ab}=\tilde\Theta_i^\T\Pi\,\tilde\Theta_i/(N-2k)$, where the
matrix $\Pi$ has, for each $t=k,\dots,N-k-1$, a one in the row of the $a$-th
entry of $\tth_i(t+j)$ and the column of the $b$-th entry of $\tth_i(t-k)${,
so that $\norm\Pi=1$ and $\norm\Pi_F^2=N-2k$}.
As the agents are independent,
$\tilde\Theta:=\col(\tilde\Theta_1,\dots,\tilde\Theta_m)$ is zero-mean Gaussian
with a block-diagonal covariance $\Sigma$, and $\norm\Sigma\le c_1$
by~\eqref{eq:cov}, so we may write $\tilde\Theta=\Sigma^{1/2}g$ with
$g\sim\mc N(0,I)$. Then
\begin{equation*}
\sum_{l=1}^m\omega_l\big(\hat M_{j,l}-M_{j,l}\big)_{ab}=g^\T Bg-\E[g^\T Bg],\qquad
B:={\tfrac{1}{N-2k}}\Sigma^{1/2}\blkdiag(\omega_1\Pi,\dots,\omega_m\Pi)\Sigma^{1/2},
\end{equation*}
with $\norm B\le c_1\norm\omega/(N-2k)$ and $\norm B_F\le c_1\norm\omega/\sqrt{N-2k}$.
By the Hanson--Wright inequality \citep[Theorem~1.1]{MR-RV:13}, for an
absolute constant $c_0>0$ and all $s\ge0$,
\begin{equation}\label{eq:hw}
\Prob\big(|g^\T Bg-\E[g^\T Bg]|>s\big)\le2\exp\Big(-c_0\min\Big\{\frac{s^2}{\norm B_F^2},\,\frac{s}{\norm B}\Big\}\Big).
\end{equation}
Let $\ell:=\log(2p^2/\delta)/c_0$ and $s:=c_1\norm\omega\sqrt{\ell/(N-2k)}$. 
{If $N-2k\ge\ell$, both ratios in~\eqref{eq:hw} are at least $\ell$, so
each of the $p^2$ entries of $\sum_l\omega_l(\hat M_{j,l}-M_{j,l})$ exceeds
$s$ in absolute value with probability at most $2e^{-c_0\ell}=\delta/p^2$,
and} a union bound gives
$\norm{\sum_l\omega_l(\hat M_{j,l}-M_{j,l})}\le ps$ with probability at least
$1-\delta$. 
Now let
\begin{equation}\label{eq:cN0}
c:=p\,c_1\sqrt{\frac{2\big(1+\log_2(2p^2)\big)}{c_0}},\qquad
N_0(\delta):=4k+\frac{1}{c_0}\log\frac{2p^2}{\delta} .
\end{equation}
If $N\ge N_0({\delta})$, 
then 
{$N-2k\ge N/2$ and $N-2k\ge\ell$, and, as $\delta\le1/2$,}
$\log(2p^2/\delta)\le(1+\log_2(2p^2))\log(1/\delta)$, so
$ps\le c\norm\omega\sqrt{\log(1/\delta)/N}$. Thus, for {$\delta\in(0,1/2]$
and} $N\ge N_0(\delta)$,
with probability at least $1-\delta$,
\begin{equation}\label{eq:conc-s}
\Big\lVert\sum_{l=1}^m\omega_l\big(\hat M_{j,l}-M_{j,l}\big)\Big\rVert\le c\,\norm\omega\sqrt{\frac{\log(1/\delta)}{N}},
\end{equation}
which is~\eqref{eq:conc}{, in the range $\delta\le1/2$, $N\ge N_0(\delta)$
in which it is used below}.

\emph{(iii) Consensus error:} By~\eqref{eq:cons},
$\hat M_j^{(i)}(\tau)=\sum_l(W^\tau)_{il}\hat M_{j,l}$. The weights
$\omega^{(i)}_l:=(W^\tau)_{il}-1/m$ {satisfy $\sum_l\omega^{(i)}_l=0$,
as $W^\tau$ is row stochastic,} and
$\norm{\omega^{(i)}}\le\norm{W^\tau-J}=\norm{(W-J)^\tau}\le\sigma^\tau$, as
$WJ=JW=J=J^2$. 
Hence
\begin{equation}\label{eq:consdec}
\hat M_j^{(i)}(\tau)-\bar M_j
=\frac1m\sum_{l=1}^m\big(\hat M_{j,l}-M_{j,l}\big)+\sum_{l=1}^m\omega^{(i)}_l\big(\hat M_{j,l}-M_{j,l}\big)+\sum_{l=1}^m\omega^{(i)}_l\big(M_{j,l}-\bar M_j\big).
\end{equation}
Let $\mc E_\delta$ be the event on which~\eqref{eq:conc-s} holds {with
$\delta/(4m)$ in place of $\delta$, for $j\in\{0,1\}$ and
$\omega\in\{\ones_m/m,\omega^{(1)},\dots,\omega^{(m)}\}$}. 
Since
$N\ge N_0(\delta/(4m))$, a union bound {over these $2(m+1)$ events} gives
$\Prob(\mc E_\delta)\ge
{1-2(m+1)\delta/(4m)\ge}
1-\delta$. On $\mc E_\delta$, the three
terms of~\eqref{eq:consdec} are at most $cm^{-1/2}\sqrt{\log(4m/\delta)/N}$,
$c\sigma^\tau\sqrt{\log(4m/\delta)/N}$, and
$\norm{\omega^{(i)}}_1D_M\le\sqrt m\,\sigma^\tau D_M$, so, by~\eqref{eq:em},
\begin{equation}\label{eq:momerr}
\big\lVert\hat M_j^{(i)}(\tau)-\bar M_j\big\rVert\le\epsm(N,\tau,\delta)\qquad\text{for all }i\in\mc V\text{ and }j\in\{0,1\},\ \text{on }\mc E_\delta .
\end{equation}

\emph{(iv) Bias of the pooled moments:} By Section~\ref{sec:alg},
$\bar M_1=A\bar M_0+\bar\Xi$. By~\eqref{eq:betabound}{, \eqref{eq:Bprime},
and the Cauchy--Schwarz inequality}, $\norm{\Xi_i}\le(1+\norm A)\,b_{k,i}{\Bp{i}}$, so
\begin{equation}\label{eq:pooled}
\norm{\bar\Xi}\le\eb:=\frac{1+\norm A}{m}\sum_{i=1}^mb_{k,i}{\Bp{i}} .
\end{equation}

\emph{{(v)} Perturbation of the IV estimator:} Since
$\bar M_0=A^k\bar\Sigma_\theta+\bar\Delta_0$, Weyl's
inequality $\sm(X+Y)\ge\sm(X)-\norm Y$, 
$\norm{\bar\Delta_0}\le\nu_k/2$ and $\epsm\le\nu_k/8$, and~\eqref{eq:momerr}
give $\sm(\bar M_0)\ge\nu_k/2$ and $\sm(\hat M_0^{(i)}(\tau))\ge3\nu_k/8$ on
$\mc E_\delta$, where $\nu_k>0$ as $A$ is invertible and
$\bar\Sigma_\theta\succ0$. On $\mc E_\delta$, {writing $\hat M_j$ for
$\hat M_j^{(i)}(\tau)$,} the identities $\bar M_1\bar M_0^{-1}=A+\bar\Xi\bar M_0^{-1}$
and $X^{-1}-Y^{-1}=-Y^{-1}(X-Y)X^{-1}$ give
\begin{equation}\label{eq:pert}
\Ai-A=\big(\hat M_1-\bar M_1\big)\hat M_0^{-1}-\bar M_1\bar M_0^{-1}\big(\hat M_0-\bar M_0\big)\hat M_0^{-1}+\bar\Xi\bar M_0^{-1},
\end{equation}
so, by~\eqref{eq:momerr} and $\norm{\bar M_1\bar M_0^{-1}}\le\norm A+2\eb/\nu_k$,
for all $i\in\mc V$ on $\mc E_\delta$,
\begin{equation}\label{eq:final-id}
\norm{\Ai-A}\le\frac{8}{3\nu_k}\Big(1+\norm A+\frac{2\eb}{\nu_k}\Big)\epsm(N,\tau,\delta)+\frac{2\eb}{\nu_k}.
\end{equation}
By~\eqref{eq:em} and $m^{-1/2}+\sigma^\tau=(1+\sqrt m\sigma^\tau)/\sqrt m$, the
first term gives the finite-sample and heterogeneity terms of
Theorem~\ref{thm:id} with
$C_A:=\frac{8c}{3\nu_k}\big(1+\norm A+\frac{2\eb}{\nu_k}\big)$ and
$C_W:=C_A/c$. The second term is the
bias residual{: by~\eqref{eq:pooled}, \eqref{eq:betabound}, and
$\frac1m\sum_i\sqrt{\trace Q_i}\le\sqrt{\trace\bar Q}$ (concavity of the
square root)},
\begin{equation}\label{eq:CB}
\begin{gathered}
\frac{2\eb}{\nu_k}\le C_B\big(k\norm{A-I}+\sqrt{k\trace\bar Q}\big),\\
C_B:=\frac{2(1+\norm A)C_\rho}{\nu_k}\max_{i\in\mc V}\big(\Gamma_i\chi_i\Bp{i}\big)\max\Big\{\frac{\max_iB_{\theta,i}}{\sqrt3},\frac1{\sqrt2}\Big\}.
\end{gathered}
\end{equation}
The constants $C_A$, $C_W$, $C_B$ do not depend on $N$, $\tau$, or $\delta$.

\paragraph{Proof of Lemma~\ref{lemma:3}:}%

Let $X(t)$, $S(t)$, $G(t)\in\R^{m\times n}$ be the matrices with $i$-th rows
$x_i(t)^\T$, $s_i(t)^\T$, and $\gamma_i(t)^\T$, where
$\gamma_i(t)=\nabla_x\hat f_i(x_i(t),t)$ for $t\ge N$, and let
$\bar g(t):=\frac1m\sum_i\gamma_i(t)$. 
For $t\ge N$ the forecasts are fixed functions of the data, and~\eqref{eq:gt}
is the deterministic recursion
\begin{equation}\label{eq:gtmat}
X(t+1)=W\big(X(t)-\alpha S(t)\big),\qquad S(t+1)=W\big(S(t)+G(t+1)-G(t)\big).
\end{equation}

\emph{(i) Tracking identity.} Multiplying~\eqref{eq:gtmat} by
$\frac1m\ones_m^\T$ gives $\bar x(t+1)=\bar x(t)-\alpha\bar s(t)$ and
$\bar s(t+1)=\bar s(t)+\bar g(t+1)-\bar g(t)$ for all $t\ge0$. Since
$\bar s(0)=\bar g(0)$, induction gives
\begin{equation}\label{eq:trackid}
\bar s(t)=\bar g(t)=\frac1m\sum_{i=1}^m\nabla_x\hat f_i(x_i(t),t)
\qquad\text{and}\qquad
\bar x(t+1)=\bar x(t)-\alpha\bar g(t),\qquad t\ge N .
\end{equation}

\emph{(ii) Iterate disagreement.} By~\eqref{eq:gtmat}, $(I-J)W=W-J$, and
$(W-J)\ones_m=0$,
so
\begin{equation}\label{eq:a}
a_{t+1}\le\sigma\,(a_t+\alpha c_t).
\end{equation}

\emph{(iii) Optimization error.} Write
$\bar g(t)=\frac1m\nabla_x\hat F(\bar x(t),t)+r_t$, where, by $L$-smoothness
and the Cauchy--Schwarz inequality,
\begin{equation*}
r_t:=\frac1m\sum_{i=1}^m\big(\nabla_x\hat f_i(x_i(t),t)-\nabla_x\hat f_i(\bar x(t),t)\big),\qquad
\norm{r_t}\le\frac Lm\sum_{i=1}^m\norm{x_i(t)-\bar x(t)}\le\frac{L}{\sqrt m}\,a_t .
\end{equation*}
By~\eqref{eq:trackid}, $\bar x(t+1)=T_t(\bar x(t))-\alpha r_t$,
and by~\eqref{eq:lip}, $\norm{\hat x^*(t+1)-\hat x^*(t)}\le L_*d_t$. Therefore
$\norm{\bar x(t+1)-\hat x^*(t+1)}\le(1-\frac{\alpha\mu}{m})\norm{\bar x(t)-\hat x^*(t)}+\alpha\norm{r_t}+L_*d_t$,
that is,
\begin{equation}\label{eq:b}
b_{t+1}\le\alpha L\,a_t+\big(1-\tfrac{\alpha\mu}{m}\big)b_t+\sqrt m\,L_*\,d_t .
\end{equation}

\emph{(iv) Tracker disagreement:} 
By~\eqref{eq:gtmat} and $(W-I)\ones_m=0$,
$X(t+1)-X(t)=(W-I)\big(X(t)-\ones_m\bar x(t)^\T\big)-\alpha WS(t)$, so
$\norm{X(t+1)-X(t)}_F\le2a_t+\alpha\norm{S(t)}_F\le2a_t+\alpha\big(c_t+\sqrt m\norm{\bar s(t)}\big)$,
where, by~\eqref{eq:trackid}, $\nabla_x\hat F(\hat x^*(t),t)=0$, and the
$mL$-smoothness of $\hat F(\cdot,t)$,
$\norm{\bar s(t)}\le\frac1m\norm{\nabla_x\hat F(\bar x(t),t)-\nabla_x\hat F(\hat x^*(t),t)}+\norm{r_t}\le\frac{L}{\sqrt m}(a_t+b_t)$.
Collecting,
\begin{equation}\label{eq:c}
c_{t+1}\le\sigma\Big(L(2+\alpha L)\,a_t+\alpha L^2\,b_t+(1+\alpha L)\,c_t+\Gamma\,d_t\Big).
\end{equation}

\emph{(v) Spectral radius:} Inequalities~\eqref{eq:a}--\eqref{eq:c} gives
$e_{t+1}\le\Lambda e_t+v\,d_t$ for $t\ge N$, with
\begin{equation}\label{eq:Lambda}
\Lambda:=\begin{bmatrix}\sigma&0&\alpha\sigma\\ \alpha L&1-\alpha\mu/m&0\\
\sigma L(2+\alpha L)&\sigma\alpha L^2&\sigma(1+\alpha L)\end{bmatrix},\qquad
v:=\begin{bmatrix}0\\ \sqrt m\,L_*\\ \sigma\Gamma\end{bmatrix},
\end{equation}
both nonnegative, since $\alpha\mu/m\le\mu^2/\big(m(mL+\mu)L\big)<1$ by the
step-size condition. 

\emph{(vi) Conclusion:} Since $\Lambda$ and $v$ are nonnegative, unrolling
the recursion from $t=N$ gives
$0\le e_t\le\Lambda^{t-N}e_N+\sum_{q=N}^{t-1}\Lambda^{t-1-q}v\,d_q$. Fix $\rgt\in(\rho(\Lambda),1)$ and
$c_\Lambda:=\sup_{j\ge0}\norm{\Lambda^j}/\rgt^{\,j}<\infty$, 
\begin{equation*}
\norm{e_t}\le c_\Lambda\Big(\rgt^{\,t-N}\norm{e_N}+\norm v\sum_{q=N}^{t-1}\rgt^{\,t-1-q}d_q\Big).
\end{equation*}
Finally,
$\norm{x_i(t)-\hat x^*(t)}\le\norm{x_i(t)-\bar x(t)}+\norm{\bar x(t)-\hat x^*(t)}\le a_t+b_t/\sqrt m\le\sqrt2\,\norm{e_t}$,
which is~\eqref{eq:gtbound} with $\cgt:=\sqrt2\,c_\Lambda\max\{1,\norm v\}$.

\paragraph{Proof of Theorem~\ref{thm:main}:}
Fix $t\ge N$ and let $H:=t-(N-k)$, so that $\hth_i(t)=(\Ai)^H\tth_i(N-k)$
by~\eqref{eq:fore}. Since $\theta(t),\hth(t)\in\Theta$, the triangle inequality
and~\eqref{eq:lip} give
\begin{equation}\label{eq:split4}
\norm{x_i(t)-x^*(t)}\le\norm{x_i(t)-\hat x^*(t)}+L_*\sum_{j=1}^m\norm{\hth_j(t)-\theta_j(t)} .
\end{equation}

\emph{(i) Powers of the identified dynamics:} On $\mc E_\delta$,
$\norm{\Ai-A}\le\eid(\delta)$ by Theorem~\ref{thm:id}. Expanding
$(\Ai)^j=\big(A+(\Ai-A)\big)^j$, each of the $\binom jl$ products with $l$
factors $\Ai-A$ has norm at most $C_\rho^{l+1}\rho_A^{\,j-l}\eid(\delta)^l$, as
$C_\rho\ge1$. Summing, and using
$(\Ai)^H-A^H=\sum_{l=0}^{H-1}(\Ai)^l(\Ai-A)A^{H-1-l}$ with
$\rho_A\le\hat\rho$, we get on $\mc E_\delta$
\begin{equation}\label{eq:powers}
\norm{(\Ai)^j}\le C_\rho\big(\rho_A+C_\rho\eid(\delta)\big)^j=C_\rho\hat\rho^{\,j},\qquad
\norm{(\Ai)^H-A^H}\le C_\rho^2H\hat\rho^{\,H-1}\eid(\delta).
\end{equation}

\emph{(ii) Term~(iii) of~\eqref{eq:main} (the forecast error):}
By~\eqref{eq:powers}
and~\eqref{eq:recon}, and the independence of the zero-mean process noises,
the three terms of~\eqref{eq:fdec} satisfy
\begin{equation*}
\begin{aligned}
\Ed\big\lVert\big((\Ai)^H-A^H\big)\tth_i(N-k)\big\rVert&\le C_\rho^2H\hat\rho^{\,H-1}\eid(\delta)\,\Bp{i},\\
\Ed\norm{A^H\tilde e_i}&\le C_\rho\rho_A^{\,H}\,\ew_i,\\
\Ed\Big\lVert\sum_{j<H}A^jw_i(t-1-j)\Big\rVert&\le\Big(\trace\sum_{j<H}A^jQ_i(A^j)^\T\Big)^{1/2}=\sigma_{H,i},
\end{aligned}
\end{equation*}
and $\sigma_{H,i}^2\le\trace\Sigma_{\theta,i}$ since
$\Sigma_{\theta,i}=\sum_{l\ge0}A^lQ_i(A^l)^\T$. Hence
$\Ed\norm{\hth_i(t)-\theta_i(t)}\le\ef_i(N,t;\delta)$ for every $i\in\mc V$,
giving term~(iii) of~\eqref{eq:main}. 

\emph{(iii) Terms~(i) and~(ii) of~\eqref{eq:main} (the tracking error):}
By~\eqref{eq:fore},
$\hth_j(q+1)-\hth_j(q)=(\hat A^{(j)})^{q-N+k}(\hat A^{(j)}-I)\tth_j(N-k)$, so,
on $\mc E_\delta$, \eqref{eq:powers} and
$\norm{\hat A^{(j)}-I}\le\norm{A-I}+\eid(\delta)$ give, for $q\ge N$,
\begin{equation*}
d_q\le\sum_{j=1}^m\norm{\hth_j(q+1)-\hth_j(q)}\le C_\rho\hat\rho^{\,q-N+k}\big(\norm{A-I}+\eid(\delta)\big)\sum_{j=1}^m\norm{\tth_j(N-k)} .
\end{equation*}
As $\rgt,\hat\rho\le\rho_\star$,
$\sum_{q=N}^{t-1}\rgt^{\,t-1-q}\hat\rho^{\,q-N+k}\le\hat\rho^{\,k}(t-N)\rho_\star^{\,t-N-1}$.
Inserting this in~\eqref{eq:gtbound}, which holds for every realization of
the data, and taking $\Ed$ with $\E\norm{\tth_j(N-k)}\le\Bp{j}$ gives
\begin{equation*}
\Ed\norm{x_i(t)-\hat x^*(t)}\le\cgt\rgt^{\,t-N}\Ed\norm{e_N}+\cgt C_\rho\hat\rho^{\,k}(t-N)\rho_\star^{\,t-N-1}U_N,
\end{equation*}
which are terms~(i) and~(ii).

\end{document}